\documentclass[conference]{IEEEtran}
\IEEEoverridecommandlockouts
\usepackage{cite}
\usepackage{amsmath,amssymb,amsfonts}
\usepackage{algorithmic}
\usepackage{graphicx}
\usepackage{textcomp}
\usepackage{xcolor}
\def\BibTeX{{\rm B\kern-.05em{\sc i\kern-.025em b}\kern-.08em
    T\kern-.1667em\lower.7ex\hbox{E}\kern-.125emX}}
\begin{document}

\title{SDVDiag: Multimodal Causal Discovery for Online Diagnosis in Software-defined Vehicles\\
}
\author{
\IEEEauthorblockN{Matthias Weiß, Athreya Hosahalli Prakash, Falk Dettinger, Nasser Jazdi and Michael Weyrich}
\IEEEauthorblockA{
\textit{Institute of Industrial Automation and Software Engineering (IAS)} \\
\textit{University of Stuttgart} \\
Pfaffenwaldring 47, 70550 Stuttgart, Germany \\
E-Mail: \{matthias.weiss, athreya.hosahalli-prakash, falk.dettinger, nasser.jazdi, michael.weyrich\}@ias.uni-stuttgart.de}}

\maketitle

\begin{abstract}
The transition toward software-defined vehicles concentrates an increasing share of vehicle functionality into distributed software services, where failures propagate through service dependencies and the surface symptom is often several causal hops away from the underlying defect. Existing approaches to causal root-cause analysis in such systems address this only partially: they typically reason over a single observability modality and operate in an offline, operator-driven mode that does not match the demands of continuous vehicle operation. This paper presents SDVDiag, a multimodal causal-discovery pipeline that fuses log-based and metric-based service representations into a shared embedding space before graph construction, coupled with an anomaly-driven trigger that converts the diagnostic platform from a manually operated batch tool into a continuously running online system. Evaluation on an Autonomous Valet Parking testbed shows that the multimodal pipeline produces sparser causal graphs than a metrics-only baseline (134 vs. 182 edges on average) and consistently outperforms it in edge-weighted reward against an expert knowledge graph at every stage of human-feedback refinement, showing a 2.4-fold improvement over the baseline after 60 feedback queries. An end-to-end fault-injection scenario further demonstrates that the integrated trigger correctly recovers a true root cause located two causal hops upstream of the observable symptom.
\end{abstract}

\begin{IEEEkeywords}
software-defined vehicles, root-cause analysis, causal discovery, multimodal learning, anomaly detection, online diagnosis
\end{IEEEkeywords}

\section{Introduction}
\label{sec:intro}
 
Modern vehicles are increasingly defined by the software they run. Driver assistance, infotainment, fleet coordination, remote diagnostics, and a growing range of comfort and convenience features are no longer realized by isolated electronic control units but by distributed software services that span the in-vehicle network and a connected backend infrastructure~\cite{guo2025automated, dettinger2024future, lu2025research}. This shift toward software-defined vehicles (SDVs)~\cite{Stuempfle2025SDV} brings substantial advantages, from over-the-air feature deployment to cross-fleet learning, but it also fundamentally changes the failure modes vehicle manufacturers must contend with. The reliability of a connected function no longer depends on a single ECU but on the joint behavior of many cooperating services, the network paths between them, and the backend resources they share~\cite{dettinger2024future}.
 
For the diagnostic infrastructure that vehicle manufacturers operate, this shift is consequential. A failure in one service rarely manifests where it originates: a memory leak in a backend service can surface as a latency spike on an unrelated in-vehicle function, and the symptoms an operator observes are often several causal hops removed from the underlying defect. Conventional diagnostic techniques in the automotive industry, which were developed around fault codes emitted at ECU granularity, do not capture this propagation structure~\cite{weiss2024simulating}. As the number of cooperating services grows, manually reconstructing causal chains through service dependencies becomes increasingly impractical, and the time required to localize a root cause begins to dominate the overall mean-time-to-repair.
 
Bringing causality explicitly into the diagnostic process is therefore an active area of research, but two practical obstacles have so far prevented it from being applied in continuous vehicle operation. First, a vehicle backend produces several types of operational data --- numerical metrics, application logs, and distributed traces --- each carrying information that the others do not, yet most existing approaches reason over only one of them at a time~\cite{wang2024aiopssurvey}. Second, vehicle backends are not static: software is updated over the air, fleets change, and usage patterns shift over time, so a model trained on past data is not guaranteed to describe current behavior. Existing methods typically assume a fixed system and rely on the operator to localize a suspicious service before the analysis begins, which is workable for retrospective investigation but does not fit the demands of continuous operation~\cite{pham2024howfar}.
 
The contribution of this paper is to address both limitations within a single platform. Concretely, we extend the existing SDVDiag framework~\cite{weiss2025sdvdiag} along two complementary directions: a multimodal causal-discovery pipeline that fuses log-based and metric-based service representations into a shared embedding space before graph construction, and an integration of anomaly detection as the trigger that converts the framework from a manually operated, batch-mode tool into a continuously running online diagnosis system. We evaluate the resulting platform on an Autonomous Valet Parking testbed, demonstrating that the multimodal pipeline produces sparser and more accurate causal graphs than a metrics-only baseline and that the integrated trigger correctly recovers the true root cause in a fault-injection scenario where the surface symptom is two causal hops removed from its origin.
 
The remainder of this paper is structured as follows. Section~\ref{sec:sota} reviews the state of the art in causal root-cause analysis and identifies the gaps the paper addresses. Section~\ref{sec:architecture} describes the proposed platform architecture, including the log-processing pipeline, the embedding fusion mechanism, and the online diagnosis loop. Section~\ref{sec:evaluation} presents the experimental evaluation. Section~\ref{sec:conclusion} concludes with a discussion of limitations and outlook on future work.
 

\section{State of the Art}
\label{sec:sota}
The transition toward software-defined vehicles concentrates an increasing share of vehicle functionality into distributed software services that span the in-vehicle network and connected backend infrastructure~\cite{Stuempfle2025SDV}. As this software stack grows, failures propagate across services through shared resources and request chains, so that the symptom an operator observes is frequently several hops away from its underlying cause. Conventional diagnostic techniques in the automotive industry, built around fault codes and ECU-level reporting, capture none of this propagation structure: they tell the operator \emph{what} has gone wrong locally, but not \emph{why}, leaving manual investigation of backend services and the reconstruction of causal chains as the only path to a root cause. For the diagnostic loads anticipated in software-defined vehicles, this is not sustainable.

A growing body of work on root-cause analysis (RCA) responds to this challenge by representing the system as a causal graph across all software modules or services and traversing it to localize faults. Established methods construct such graphs from a single data modality. Metrics-based approaches such as CIRCA~\cite{li2022circa} and DejaVu~\cite{li2022dejavu} infer dependencies from time-series telemetry, while LogBERT~\cite{guo2021logbert} and similar log-based methods extract anomaly signal from unstructured event streams. Each modality on its own captures only one facet of system behavior: metrics expose how heavily a service is loaded but say little about why, while logs describe internal events but lack the continuous quantitative signal needed to compare services. The integration of multiple modalities into a single causal-discovery pipeline is consequently identified by recent surveys as an open challenge~\cite{wang2024aiopssurvey,pham2024howfar}, and initial proposals such as DeepTraLog~\cite{zhang2022deeptralog}, Eadro~\cite{lee2023eadro}, and MULAN~\cite{zheng2024mulan} demonstrate that fused representations yield more accurate causal graphs than any single-modality variant.

A second limitation cuts across both single-modality and multimodal proposals: nearly all are offline by design. They
assume that the dependency structure is fixed at analysis time, that representative training data is available in
advance, and that the resulting graph is then applied retrospectively to incidents already collected~\cite{pham2024howfar}. Vehicle backends subject to over-the-air updates, varying operational profiles, and changing fleet composition violate the first two assumptions in normal operation~\cite{Stuempfle2025SDV}, and the third assumption only holds when an operator manually selects which incident to analyze and over what window. The latter is itself a significant constraint: in continuous operation the operator must first localize a suspicious service before the causal-discovery pipeline can be applied to it, which both delays diagnosis and makes the localization quality dependent on whichever symptom happens to be most visible~\cite{weiss2023continuous}. An anomaly detector running continuously over the same telemetry is a natural candidate to provide that localization automatically, with the additional benefit that the most-anomalous service is, by construction, a more informative entry point for causal traversal than an arbitrarily-chosen one \cite{wang2024aiopssurvey, weiss2024review}.
 
As such, two fundamental gaps remain for diagnosis in software-defined vehicles: causal-discovery pipelines remain largely single-modality, and they remain offline by design, requiring representative training data and a fixed dependency structure at analysis time. Closing both gaps simultaneously is the goal of this paper.


\section{Platform Architecture}
\label{sec:architecture}

\begin{figure*}[t!b]
    \centering
    \includegraphics[width=\linewidth]{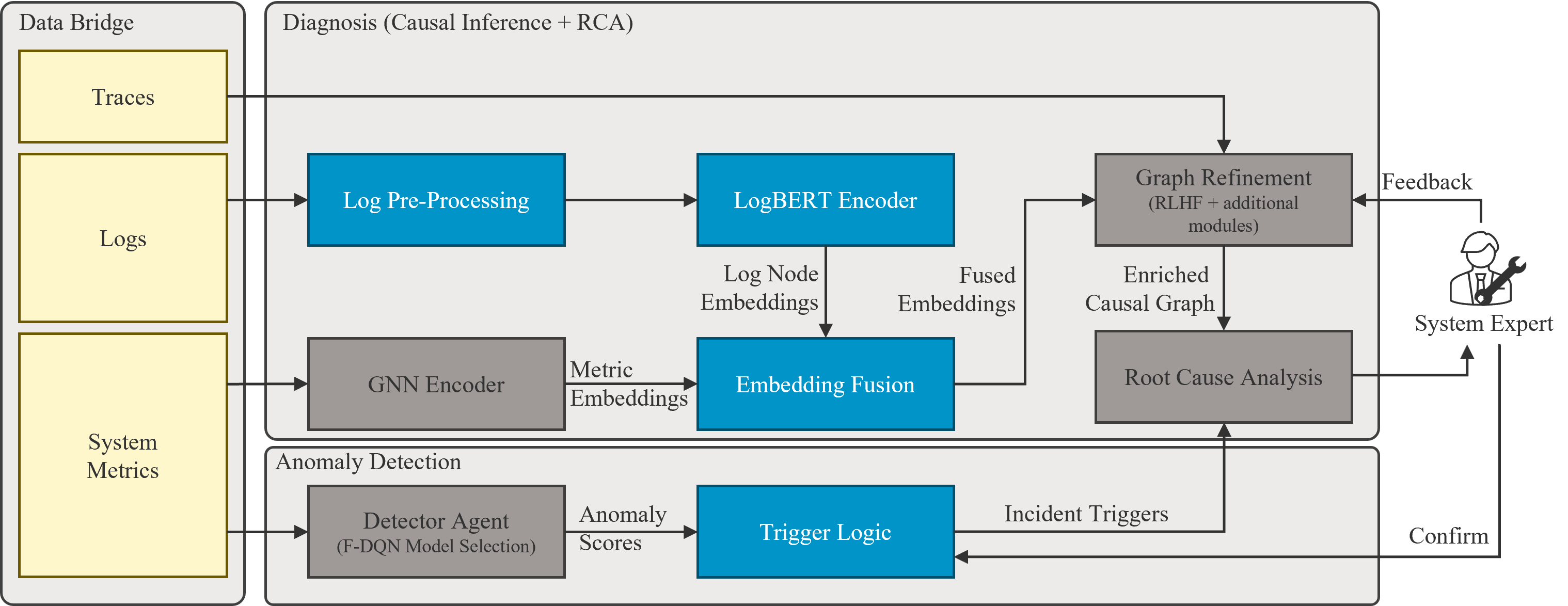}
    \caption{Overview of the full multimodal pipeline. Blue boxes represent new modules added by this work. Logs are aggregated, processed and encoded, followed by a fusion with the existing metric embeddings. This is concluded by the graph reconstruction and the actual root cause analysis, which is triggered by the anomaly detection.}
    \label{fig:pipeline}
\end{figure*}

The proposed diagnostic pipeline is implemented as an extension of \emph{SDVDiag}, a modular diagnosis platform for connected-vehicle functions introduced in prior work~\cite{weiss2025sdvdiag}. SDVDiag organizes diagnostic capability into four functional layers: a \emph{data aggregation layer} that collects telemetry from the vehicle fleet, a \emph{graph creation} layer that derives a causal graph from the telemetry and given context, an \emph{incident analysis} layer that localizes root causes once an anomaly is suspected, and a \emph{learning environment} that supports continuous refinement of the encoders and policies. This work targets two areas: the graph creation, where a log-based modality and a fusion mechanism is introduced, and the boundary between incident analysis and the anomaly detection, where the loop between detection and diagnosis is closed and enabled for online operation.

Of the components reused from prior work, the graph creation layer uses a Graph Neural Network (GNN) to encode metrics into node embeddings that are subsequently refined by a Reinforcement Learning from Human Feedback (RLHF) loop~\cite{wang2023hrlhf} with trace-based pruning, while the anomaly detection subsystem produces per-service anomaly scores via a model-selection strategy. The remaining components, including the random-walk root-cause ranking, are reused unchanged; full details on the existing implementation are available in~\cite{weiss2026context, weiss2026adselection}.

Figure~\ref{fig:pipeline} shows the overall structure of the proposed pipeline. Raw telemetry is streamed from the monitoring stack through a Kafka-based data bridge that publishes metrics, logs, pod metadata, and traces on dedicated topics, decoupling the analysis path from the availability of Prometheus and Loki at query time and, critically, providing deterministic time alignment between metric and log windows. On the causal-discovery path, the metric stream is consumed by the existing GNN encoder, while the log stream is processed by the new log-processing pipeline and encoded with a LogBERT-based model. The two modality-specific embeddings are aligned and combined by a fusion autoencoder; the resulting joint embedding is used to reconstruct a base causal graph via similarity scoring. This graph is then refined iteratively by the RLHF loop, producing the enriched causal graph on which random-walk root-cause ranking operates. In parallel, the anomaly detector consumes the metric stream and produces per-service anomaly scores; a threshold logic translates sustained anomalies into automated triggers for the incident-analysis workflow. The three additions -- log processing, embedding fusion, and the online trigger -- are described in the following subsections.

\subsection{Log Processing}
\label{sec:log-processing}

Application logs carry semantic information about service state that is not recoverable from numerical metrics alone. A memory leak, for instance, may manifest both as a rising memory-utilization curve and as a sequence of warnings about allocation failures; only the latter identifies the specific failure mode. Integrating logs into a causal-discovery pipeline is non-trivial, however: logs are unstructured, high-dimensional, noisy, and produced at rates that can easily exceed metric volume by two orders of magnitude. A direct ingestion of raw log lines would dominate both memory and computation during embedding generation, and would be susceptible to the high cardinality of the template space. The proposed log-processing pipeline addresses these concerns in two stages. The first stage compresses raw logs into a semantically meaningful, length-bounded representation suitable for learning. The second stage, implemented via the LogBERT encoder, transforms these representations into fixed-size service-level node embeddings.

The compression stage proceeds as follows. Raw log lines are first \emph{normalized and masked}: timestamps, UUIDs, IP addresses, and other variable tokens that might contain sensitive user information are replaced by placeholders, removing entropy that does not contribute to causal structure. The sanitized lines are then passed to \emph{Drain3}~\cite{he2017drain}, an online parser that clusters log entries into structured templates separating constant message patterns from parameters. Templated logs are grouped by service into fixed-width time windows (e.g., 300\,s, determined by the configured analysis cadence). Within each window, consecutive occurrences of the same template are collapsed by Run-Length Encoding (RLE) into quadruplets of the form $(\Delta t, \texttt{service}, \texttt{template}, \texttt{count})$, where $\Delta t$ is the time offset from the window start. This compression preserves the order of distinct events while discarding repetition, which is the dominant source of redundancy in service logs.


Before LogBERT can consume these quadruplets, the \texttt{template} field must itself be projected into vector space. A hash-based identifier would be trivial to compute but would treat semantically similar templates as unrelated, losing all generalization to log variations introduced by code changes. We therefore embed templates with the \texttt{all-MiniLM-L6-v2} sentence transformer~\cite{transformers2025minilm}, which produces 384-dimensional vectors and is trained for short-sequence semantic similarity that is commonly present in log files. Each quadruplet is finally composed into a single 384-dimensional event vector as
\begin{equation}
\mathbf{e} = W_T\,\mathbf{v}_{\text{tmpl}} + E_S(s) + W_C\,\log(1{+}c) + W_{\Delta}\,(\Delta t / \tau),
\label{eq:event-vector}
\end{equation}
where $\mathbf{v}_{\text{tmpl}}$ is the sentence-transformer template embedding, $E_S(s)$ is a learnable service-identifier embedding, $c$ is the RLE count with logarithmic scaling to compress its dynamic range, and $\Delta t / \tau$ is the normalized time offset. The weights $W_T$, $W_C$, and $W_\Delta$ are linear projections into the 384-dimensional event-vector space.

The \emph{LogBERT} encoder~\cite{guo2021logbert} transforms the RLE event sequence of each service window into a fixed-size log-based node embedding. Recurrent alternatives process sequences in a strictly forward direction, which is inadequate when the goal is a holistic window-level summary: the contextual embedding of an event should depend on both its predecessors and its successors within the window. LogBERT's bidirectional self-attention, applied over a stack of Transformer encoder layers with multi-head attention (8 heads, feedforward dimension 1024, GELU activation, residual connections, and layer normalization), produces per-event contextual representations that are informed by the entire window. A weighted attention-pooling layer then aggregates the per-event representations into a single service-level node embedding per window.

The encoder is pre-trained in a self-supervised manner via \textbf{Masked Event Modeling}, a paradigm analogous to masked language modeling but adapted for structured log event sequences. During pre-training on normal operational logs, a random subset of input event vectors is masked, and the model is trained under a \textbf{multi-task learning} objective to reconstruct properties of the masked events through three auxiliary prediction heads attached to the final Transformer layer. Each head targets a distinct semantic facet of the masked event: template reconstruction under a combined MSE and cosine similarity loss, service identification under cross-entropy, and repetition count estimation under MSE. The total loss is a weighted sum of these three objectives,
\begin{equation}
    \mathcal{L}_{\text{total}} = \lambda_1 \mathcal{L}_{\text{template}} + \lambda_2 \mathcal{L}_{\text{service}} + \lambda_3 \mathcal{L}_{\text{count}},
\end{equation}
where $\lambda_1, \lambda_2, \lambda_3$ are scalar weight factors that balance the relative contribution of each task, jointly driving the encoder to capture message content, service topology, and event frequency within a unified representation. Crucially, because the template input is a dense semantic embedding rather than a discrete one-hot template index, log templates unseen during pre-training are still meaningfully represented through their proximity to known templates in the embedding space, eliminating the retraining cycle that index-based encoders require whenever new templates emerge.

\begin{figure}[t!b]
    \centering
    \includegraphics[width=\columnwidth]{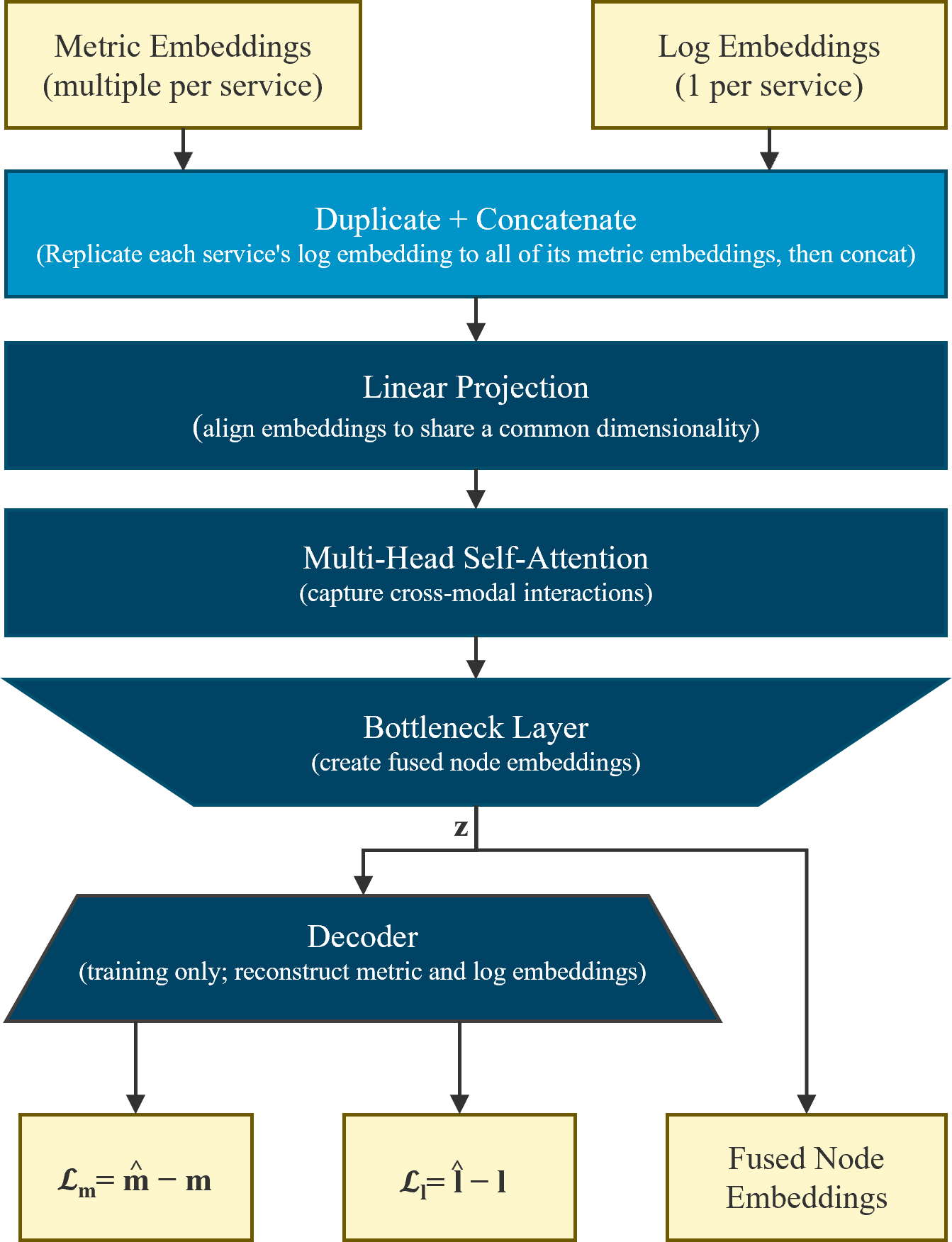}
    \caption{Fusion of metrics and logs. An encoder is trained to generate fused embeddings by a combined reconstruction loss.}
    \label{fig:fusion}
\end{figure}

\subsection{Embedding Fusion}
\label{sec:fusion}
The GNN metric encoder emits one node embedding per service-metric pair (e.g., \texttt{valetparking\_cpu\_usage}, \texttt{valetparking\_ram\_usage}), whereas LogBERT produces a single embedding per service. The two representations also occupy different latent spaces with different scales, training objectives, and dimensionalities. A direct concatenation would therefore not yield a space on which the cosine-similarity-based RLHF Policy is well-defined. As such, this approach adopts a \emph{pre-fusion} strategy that aligns the two modalities into a shared latent space before the causal graph is constructed. A \emph{post-fusion} design, i.e., building independent causal graphs per modality and combining them at the graph level, is architecturally modular but suffers from well-documented semantic loss: cross-modal dependencies are ignored until the final combination stage, so the per-modality graphs are each built on a narrow view of the system. Pre-fusion, by contrast, increases the input dimensionality of the fusion module but captures cross-modal correlations more effectively~\cite{yun2026contextual}. Furthermore, it has been observed that the RLHF refinement loop is more stable under a coherent global representation rather than isolated feature spaces, yielding an overall better performance of the full platform.

Figure~\ref{fig:fusion} shows the fusion autoencoder. The encoder accepts the concatenated per-node metric and log embeddings and first aligns them via a linear projection into a common dimensionality. Since the causal graph is built at service-metric granularity, the log embedding of each service is duplicated to each of that service's metric nodes. A multi-head self-attention block then captures intra- and inter-modal interactions such as the co-occurrence of a metric deviation with a characteristic log signature before a bottleneck compresses the representation into a fused latent vector $\mathbf{z}$. A symmetric decoder reconstructs both the metric and the log embedding spaces from $\mathbf{z}$, and the training objective is the sum of the two reconstruction errors,
\begin{equation}
\mathcal{L}_{\text{fusion}} = \lambda_m\,\lVert \hat{\mathbf{m}} - \mathbf{m} \rVert^2 + \lambda_l\,\lVert \hat{\mathbf{l}} - \mathbf{l} \rVert^2,
\label{eq:fusion-loss}
\end{equation}
with $\mathbf{m}, \mathbf{l}$ denoting the metric and log inputs and $\lambda_m, \lambda_l$ scalar weights balancing the two modalities during training. The dual-reconstruction objective forces $\mathbf{z}$ to retain sufficient information about both modalities rather than collapsing into the more easily reconstructed one, which yields fused embeddings that isolate causal features across modalities rather than spurious correlations.

\begin{figure*}[!tb]
    \centering
    \includegraphics[width=0.9\linewidth]{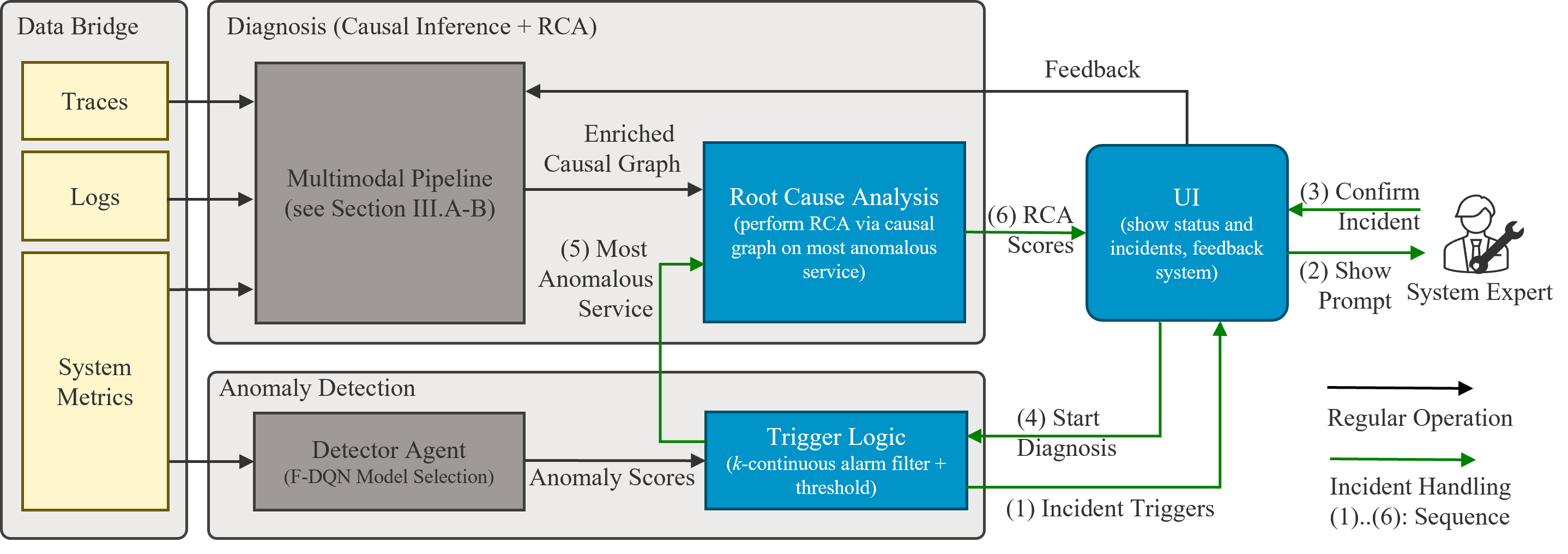}
    \caption{SDVDiag in online operation. Green edges represent the communication flow once an incident is detected.}
    \label{fig:online-loop}
\end{figure*}

The fused embeddings are consumed directly by the existing RLHF pipeline. Edge confidences are computed as cosine similarities between pairs of fused node embeddings and thresholded to produce the base causal graph. From this point onward, the refinement, trace-based pruning, domain-knowledge enrichment, and random-walk root-cause ranking proceed as in prior SDVDiag versions.

\subsection{Online Diagnosis Loop}
\label{sec:online-loop}

The preceding two subsections extend the causal-discovery path with richer node embeddings; this subsection closes the loop on the other side by integrating the anomaly detector as the automated trigger for causal analysis. In the prior SDVDiag workflow, the system engineer had to manually identify a suspicious service before the RLHF loop would begin analyzing it. This placed a cognitive burden on the operator and, in continuous operation, introduced a bottleneck that prevented the platform from functioning as a truly online diagnostic tool. The integration described here replaces the manual trigger with an anomaly-driven one.

This trigger is implemented as follows: The service that carries the highest sustained anomaly score is passed to the root-cause analysis as the target node; random walk then ranks its service-metric nodes, and the result surfaces on the unified user interface. Two mechanisms reduce false triggering. First, a \emph{$k$-continuous alarm} filter requires the anomaly score to exceed the detection threshold for at least five seconds before the service is considered suspicious, which suppresses transient score spikes. Second, a confirmation prompt is raised on the user interface; causal analysis is only invoked after the engineer confirms the alarm. The confirmation step can be disabled for fully automatic operation, but the default interactive mode has proven more useful during evaluation because it lets the operator validate the sustained-anomaly claim before committing the computational cost of a refinement cycle.

Figure~\ref{fig:online-loop} shows the end-to-end flow. Two paths run continuously alongside each other. On the analysis path, time-aligned metric and log windows from the data bridge feed the multimodal causal-discovery pipeline of Sections~\ref{sec:log-processing} and~\ref{sec:fusion}, which operates on a rolling window of telemetry (typically the last 10--15 minutes, matching the analysis batch cadence) and maintains a current \emph{enriched causal graph} at all times. The pipeline additionally consumes the trace graph from the data bridge for the trace-based pruning step. On the detection path, the anomaly detector consumes the aligned metric windows and emits per-service anomaly scores to the trigger logic.

The numbered green arrows in Figure~\ref{fig:online-loop} show the incident-handling sequence layered on top of this steady-state operation. Once the trigger logic identifies that a service exceeds the anomaly threshold for at least the configured sustained-alarm interval, it (1) raises an incident trigger to the unified user interface, which (2) prompts the system expert with a notification that an incident has been detected and offers to start the diagnosis. Upon (3) operator confirmation, the user interface (4) signals the trigger logic to release the diagnosis, which (5) forwards the most anomalous service as the target node to the root-cause analysis. The random walk traverses the current enriched causal graph and (6) returns a ranking of root-cause candidates back to the user interface, where the result is displayed alongside the live metric and anomaly views. Because the diagnosis runs against the graph that the multimodal pipeline has just produced rather than a historical snapshot, the resulting ranking reflects causal patterns specific to the anomaly under diagnosis. Aside from the incident-handling flow, the user interface also accepts RLHF edge-preference triplets from the operator, which are used by the causal refinement policies during their next iteration.

\section{Evaluation}
\label{sec:evaluation}
 
\begin{figure}[tb]
    \centering
    \includegraphics[width=\columnwidth]{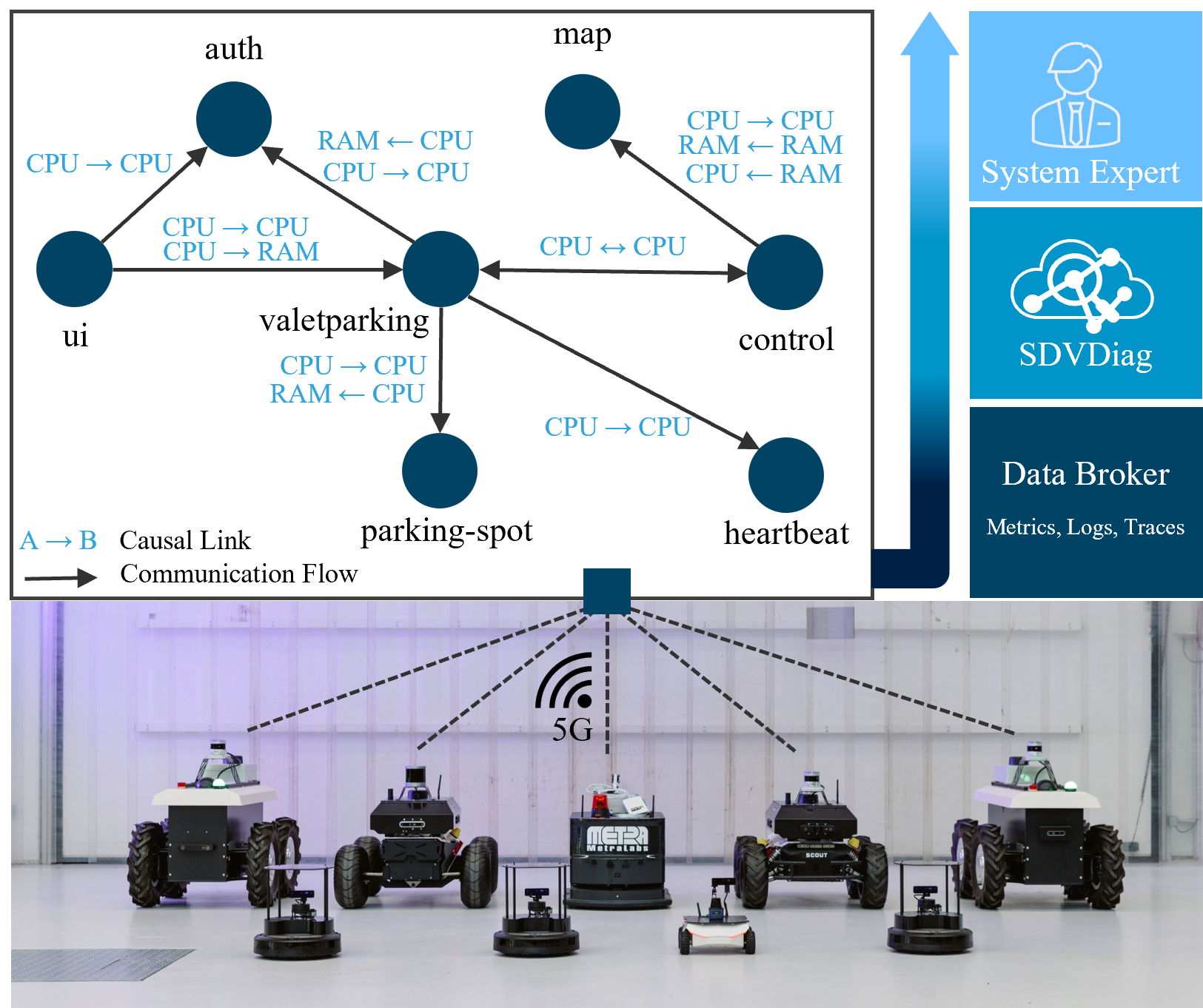}
    \caption{Evaluation testbed, including an autonomous UGV fleet and a valet parking application that are connected to SDVDiag for diagnosis.}
    \label{fig:testbed}
\end{figure}

To measure the capabilities of the devised multimodal approach in a practical scenario, the evaluation is conducted on an Autonomous Valet Parking (AVP) testbed, a representative software-defined vehicle fleet function previously established at the University of Stuttgart, which is shown in Figure~\ref{fig:testbed}. The fleet consists of different UGVs operating in a 5G test track, with \emph{park} and \emph{retrieve} workflows generating realistic telemetry load. The backend consists of seven microservices which communicate via REST APIs to handle requests by users for automated passenger pick-up and drop-off at designated locations. Metrics, logs, and distributed traces are collected via OpenTelemetry instrumentation and aggregated by Prometheus, Loki, and Zipkin respectively. The Kafka data bridge introduced in Section~\ref{sec:architecture} streams all data sources into epoch-aligned 300\,s windows, which are consumed by the multimodal pipeline and the anomaly detector in parallel.

Validating a multimodal causal-discovery pipeline is inherently more involved than validating a standard supervised model: the ground truth is a causal graph over service-metric dependencies, which is infeasible to enumerate exhaustively for a non-trivial microservice landscape~\cite{pham2024howfar}. The evaluation is therefore structured into three stages, each isolating a distinct claim. The first stage validates the log pre-processing pipeline as a complexity-reduction step that does not discard relevant signals. The second stage validates the multimodal embedding by comparing the causal graphs produced against a metrics-only baseline, using two complementary measures: graph sparsity and an edge-weighted reward derived from an expert knowledge graph. The third stage validates the integrated online-diagnosis loop on a realistic fault-injection scenario. Throughout, the baseline of comparison is the metrics-only variant of the causality mining pipeline described in~\cite{weiss2026context}. The RLHF policies themselves have been evaluated in~\cite{weiss2026context} and are not re-validated here.
 
\subsection{Log Pre-processing}
\label{sec:eval-preprocessing}
 
The log pre-processing pipeline is a prerequisite for the rest of the multimodal path: if the pipeline fails to compress logs substantially, downstream memory and compute costs become prohibitive; if it compresses too aggressively, semantic signal is lost and the fused embeddings become impoverished. The claim evaluated here is that Run-Length Encoding (RLE) combined with template-based clustering achieves substantial temporal compression without losing distinct-event information.
 
Ten disjoint batches of operational logs were collected from the AVP backend under mixed workload and passed through the full pre-processing pipeline described in Section~\ref{sec:log-processing}. For each batch, the number of raw log lines and the number of RLE quadruplets produced were recorded. Table~\ref{tab:rle} reports the results. Across all ten batches, the pipeline consistently reduces the entry count by 60--62\,\%, with a mean reduction of 61.3\,\% and a standard deviation below 0.8 percentage points, indicating that the compression ratio is stable rather than workload-dependent. Because RLE only collapses consecutive repetitions of the same template, no distinct log event is discarded; the compression operates purely on redundancy. Combined with the template-based clustering in Drain3, which resolves the high cardinality of raw log text, the pipeline transforms an unbounded, high-rate token stream into a fixed-size sequence of semantically embedded quadruplets suitable for the LogBERT encoder.
 
\begin{table}[t]
\caption{Log pre-processing: raw log lines vs.\ RLE quadruplets over 10 batches}
\label{tab:rle}
\centering
\begin{tabular}{rrrr}
\hline
Batch & Raw logs & RLE length & Reduction \\
\hline
1  & 45\,057 & 16\,953 & 62.37\,\% \\
2  & 44\,584 & 16\,744 & 62.44\,\% \\
3  & 42\,408 & 16\,401 & 61.33\,\% \\
4  & 42\,249 & 16\,592 & 60.73\,\% \\
5  & 42\,098 & 16\,062 & 61.85\,\% \\
6  & 41\,511 & 16\,568 & 60.09\,\% \\
7  & 41\,727 & 16\,307 & 60.92\,\% \\
8  & 41\,132 & 15\,912 & 61.31\,\% \\
9  & 40\,741 & 16\,015 & 60.69\,\% \\
10 & 44\,599 & 17\,394 & 61.00\,\% \\
\hline
\textbf{Mean} & \textbf{42\,611} & \textbf{16\,495} & \textbf{61.3\,\%} \\
\hline
\end{tabular}
\end{table}
 
\subsection{Multimodal Embedding and Edge Accuracy}
\label{sec:eval-embedding}
 
The central hypothesis of the paper is that fusing metric and log embeddings yields a base causal graph that more faithfully represents the true causal structure of the system than a metrics-only embedding. This hypothesis decomposes into two sub-claims. First, the fused embeddings should produce a \emph{sparser} base graph, since a richer per-node representation provides a tighter discriminator for the similarity-based edge scoring and therefore suppresses spurious correlations. Second, the edges retained in the sparser graph should be \emph{more accurate}, i.e., they should correspond more closely to dependencies a domain expert would identify as real.
 
\textbf{Graph sparsity.} Both pipelines were run on identical 300\,s telemetry windows. Averaged across ten independent windows, the metrics-only pipeline produced approximately 182 edges per base graph, while the multimodal pipeline produced approximately 134 edges. A lower edge count under the same input budget is consistent with the hypothesis that the log modality contributes discriminative information that allows the graph reconstruction to reject weakly-supported edges.
 
\textbf{Edge accuracy via expert knowledge graph.} Graph sparsity alone does not guarantee quality; a pipeline that aggressively prunes could easily produce sparse but inaccurate graphs. To ground the comparison, expert knowledge was used to estimate the causal relationships for the AVP backend (see Figure~\ref{fig:testbed}; the blue edge labels denote the causal links between metrics). These edges capture the causal dependencies that are known with high confidence from the service architecture: for example, the map server memory usage influences the control service CPU via map deserialization.
 
Each candidate causal graph $G$ produced by the pipeline is scored against the expert estimates $K$ using a reward function that evaluates both edge presence and edge direction:
 
\begin{itemize}
    \item An edge present in both $G$ and $K$ with correct direction contributes its confidence value to the total reward.
    \item An edge present in $K$ but missing from $G$ incurs a penalty of $-1$.
    \item An edge present in both but with reversed direction contributes the negative of its confidence.
    \item An edge present in $G$ but not in $K$ is ignored, since it may represent a hidden causality that the expert graph does not capture.
\end{itemize}
 
The total reward is subsequently normalized by the edge count of $G$, penalizing pipelines that achieve high recall through inflated graphs.
 
Table~\ref{tab:edge-reward} reports the edge-weighted reward and final edge count for both pipelines at five checkpoints across a human-feedback session, parameterized by the number of feedback queries answered by the domain expert. Two observations stand out. First, the multimodal pipeline outperforms the metrics-only pipeline at every checkpoint, including the zero-feedback baseline. This indicates that the improvement does not depend on feedback volume but is already visible at initialization. Second, the gap persists as feedback accumulates: at 60 feedback queries, the multimodal pipeline reaches an edge-weighted reward of 0.651 with 56 edges, compared with 0.271 and 92 edges for the metrics-only pipeline. The compounding behaviour in the absolute reward suggests that the RLHF refinement loop is more effective when operating on a multimodal base graph, consistent with the stability argument made in Section~\ref{sec:fusion}.
 
\begin{table}[t]
\caption{Edge-weighted reward vs.\ expert knowledge graph: metrics-only baseline~\cite{weiss2026context} vs.\ proposed multimodal pipeline}
\label{tab:edge-reward}
\centering
\begin{tabular}{c|rr|rr}
\hline
Feedback & \multicolumn{2}{c|}{Metrics only~\cite{weiss2026context}} & \multicolumn{2}{c}{Multimodal (ours)} \\
queries & Edges & Reward & Edges & Reward \\
\hline
0  & 182 & 0.016 & 134 & 0.057 \\
15 & 182 & 0.081 & 112 & 0.096 \\
30 & 134 & 0.059 &  75 & 0.133 \\
45 & 104 & 0.105 &  60 & 0.335 \\
60 &  92 & 0.271 &  56 & 0.651 \\
\hline
\end{tabular}
\end{table}
 
\subsection{End-to-End Online Diagnosis}
\label{sec:eval-diag}
 
The final stage evaluates the fully-integrated platform on a realistic fault scenario in which the surface symptom is spatially and causally distant from the true origin. The goal is to validate that the anomaly trigger, the multimodal causal graph, and the random-walk root-cause ranking cooperate to recover the actual root cause rather than merely flagging the most visible anomaly.
 
\textbf{Fault injection.} An oversized map payload is injected into the \texttt{map} service. The \texttt{control} service periodically retrieves and deserializes the current map, so the oversized payload drives a spike in \texttt{control} service's CPU and RAM usage. The consequent delay in request completion callbacks from \texttt{control} to \texttt{valetparking} causes park and retrieve requests to accumulate inside the \texttt{valetparking} service, producing a \emph{secondary} memory spike there. The observable symptom is therefore rising RAM usage on \texttt{valetparking}, a service that is architecturally unrelated to map management, making this a canonical case where single-service investigation would lead the operator down the wrong path.
 
\textbf{Observed behavior.} The anomaly detection detects the RAM anomaly on \texttt{valetparking} within seconds of injection. After the $k$-continuous alarm filter confirms that the anomaly persists for more than five seconds, the unified user interface raises a confirmation prompt. Upon operator confirmation, the root-cause analysis is invoked on the current enriched causal graph with \texttt{valetparking} as the target service. The random walk then traverses the causal graph upstream through the \texttt{valetparking $\to$ control} and \texttt{control $\to$ map} edges (both of which are captured in the multimodal base graph but absent or low-confidence in the metrics-only baseline), producing a ranking in which the RAM usage of the \texttt{map} service scores highest among the non-target services. The operator is thus guided to the true root cause despite the anomaly surfacing two hops downstream.
 
This scenario exercises every component of the proposed platform in sequence: the anomaly detector as trigger (Section~\ref{sec:online-loop}), the multimodal causal graph as substrate (Sections~\ref{sec:log-processing} and~\ref{sec:fusion}), and the unified interface as the control surface. A metrics-only baseline run on the same fault trace fails to produce the \texttt{control $\to$ map} service edge with sufficient confidence for the random walk to traverse, reflecting the sparser and less accurate baseline graphs observed in Section~\ref{sec:eval-embedding}.

 
\section{Conclusion}
\label{sec:conclusion}

This paper presented \emph{SDVDiag}, a multimodal causal-discovery pipeline coupled with an anomaly-driven trigger mechanism that converts an existing offline diagnostic platform into an online diagnosis system for software-defined vehicles. The technical contributions are a log-processing pipeline that compresses unstructured log streams by approximately 61\% via Run-Length Encoding while preserving semantic content, a LogBERT-based encoder that produces context-aware service-level node embeddings from the compressed log streams, and a fusion autoencoder that aligns log and metric embeddings into a shared latent space before causal-graph construction. The anomaly detector is integrated as an automated trigger for the diagnostic workflow rather than as an additional data source. Evaluation on an Autonomous Valet Parking testbed showed that the multimodal pipeline produces sparser causal graphs than a metrics-only baseline (134 vs.\ 182 edges on average) and achieves substantially higher edge-weighted reward against expert estimates at every stage of the human-feedback session, with the gap widening as feedback accumulates. The end-to-end scenario validated that the integrated platform recovers a true root cause two hops upstream of the surface anomaly, a case where a metrics-only baseline fails.
 
While significant improvements to existing approaches were observed, some limitations to this work remain and invite further investigation. The trigger-based integration relies on the anomaly detector to surface incidents and to nominate a target service for analysis, so the quality of the diagnostic workflow is bounded by the quality of detection upstream. The seven-microservice testbed used here is sufficient to validate the design and the cooperation of its components, but a larger backend would help characterize how the LogBERT encoder, the fusion autoencoder, and the RLHF refinement loop scale with node count. Likewise, the end-to-end scenario, while non-trivial, is a single fault trace; broader cross-scenario behavior is plausible based on prior work on the underlying causality mining pipeline~\cite{weiss2026context}, but a systematic benchmark across diverse fault types would be needed to confirm it.
 
Future work proceeds along two directions. The structured causal graphs produced by the platform are a natural input for a large-language-model agent acting as a higher-level diagnostic assistant, combining the interpretability of an explicit graph with the reasoning capabilities such agents provide. Beyond this, deeper integration of the anomaly signal into the causal-discovery process, specifically the weighting of edge confidences in the neighborhood of anomalous services, could improve root-cause-analysis quality for large-scale systems without compromising the trace-based pruning guarantees the platform currently relies on.

\section*{Acknowledgment}
The authors disclose that generative AI has been used to improve grammar and language of this publication. The authors have reviewed and edited all content as needed and take full responsibility for its scientific integrity and authenticity.


\bibliographystyle{IEEEtran}
\bibliography{bib}

@INPROCEEDINGS{dettinger2024future,
  author={Dettinger, Falk and Weiß, Matthias and Weyrich, Michael},
  booktitle={2024 IEEE 100th Vehicular Technology Conference (VTC2024-Fall)}, 
  title={Future Use Cases for Vehicular Communication based on Connected Functions}, 
  year={2024},
  volume={},
  number={},
  pages={1-5},
  doi={10.1109/VTC2024-Fall63153.2024.10757637 }}

@INPROCEEDINGS{weiss2025sdvdiag,
  author={Weiß, Matthias and Dettinger, Falk and Weyrich, Michael},
  booktitle={2025 IEEE International Automated Vehicle Validation Conference (IAVVC)}, 
  title={SDVDiag: A Modular Platform for the Diagnosis of Connected Vehicle Functions}, 
  year={2025},
  volume={},
  number={},
  pages={1-7},
  doi={10.1109/IAVVC61942.2025.11219581}}

@ARTICLE{Stuempfle2025SDV,
  author={Stümpfle, Johannes and Sigel, Johannes and Weiß, Matthias and Gül, Baran Can and Dettinger, Falk and Jazdi, Nasser and Hoßfeld, Max and Weyrich, Michael},
  journal={IEEE Engineering Management Review}, 
  title={The Software-Defined Vehicle: A Comprehensive Study on Current Trends and Challenges}, 
  year={2025},
  volume={},
  number={},
  pages={1-15},
  doi={10.1109/EMR.2025.3636574}
}

@INPROCEEDINGS{weiss2023continuous,
  author={Weiß, Matthias and Müller, Manuel and Dettinger, Falk and Jazdi, Nasser and Weyrich, Michael},
  booktitle={2023 IEEE 28th International Conference on Emerging Technologies and Factory Automation (ETFA)}, 
  title={Continuous Analysis and Optimization of Vehicle Software Updates using the Intelligent Digital Twin}, 
  year={2023},
  volume={},
  number={},
  pages={1-7},
  doi={10.1109/ETFA54631.2023.10275489}
}

@INPROCEEDINGS{weiss2024review,
  author={Weiß, Matthias and Thich, Stefan and Artelt, Maurice and Weyrich, Michael},
  booktitle={IECON 2024 - 50th Annual Conference of the IEEE Industrial Electronics Society}, 
  title={A survey about self-adaptive anomaly-detection in software-defined systems}, 
  year={2024},
  volume={},
  number={},
  pages={1-4},
  doi={10.1109/IECON55916.2024.10905330}
}

@inproceedings{wang2023hrlhf,
author = {Wang, Lu and Zhang, Chaoyun and Ding, Ruomeng and Xu, Yong and Chen, Qihang and Zou, Wentao and Chen, Qingjun and Zhang, Meng and Gao, Xuedong and Fan, Hao and Rajmohan, Saravan and Lin, Qingwei and Zhang, Dongmei},
title = {Root Cause Analysis for Microservice Systems via Hierarchical Reinforcement Learning from Human Feedback},
year = {2023},
isbn = {9798400701030},
publisher = {Association for Computing Machinery},
address = {New York, NY, USA},
url = {https://doi.org/10.1145/3580305.3599934},
doi = {10.1145/3580305.3599934},
booktitle = {Proceedings of the 29th ACM SIGKDD Conference on Knowledge Discovery and Data Mining},
pages = {5116–5125},
numpages = {10},
location = {Long Beach, CA, USA},
series = {KDD '23}
}

@misc{weiss2026context,
      title={SDVDiag: Using Context-Aware Causality Mining for the Diagnosis of Connected Vehicle Functions}, 
      author={Matthias Weiß and Falk Dettinger and Elias Detrois and Nasser Jazdi and Michael Weyrich},
      year={2026},
      eprint={2604.03391},
      archivePrefix={arXiv},
      primaryClass={cs.SE},
      url={https://arxiv.org/abs/2604.03391}, 
}

@ARTICLE{yun2026contextual,
  author={Yun, Sanggeon and Masukawa, Ryozo and Hassan, Raheeb and Na, Minhyoung and Imani, Mohsen},
  journal={IEEE Access}, 
  title={Contextual Fusion Strategies for Multimodal GNN-Based Reasoning: Performance and Computational Trade-Offs}, 
  year={2026},
  volume={14},
  number={},
  pages={13702-13711},
  doi={10.1109/ACCESS.2026.3653660}}

@misc{guo2021logbert,
      title={LogBERT: Log Anomaly Detection via BERT}, 
      author={Haixuan Guo and Shuhan Yuan and Xintao Wu},
      year={2021},
      eprint={2103.04475},
      archivePrefix={arXiv},
      primaryClass={cs.CR},
      url={https://arxiv.org/abs/2103.04475}, 
}

@inproceedings{pham2024howfar,
author = {Pham, Luan and Ha, Huong and Zhang, Hongyu},
title = {Root Cause Analysis for Microservice System based on Causal Inference: How Far Are We?},
year = {2024},
isbn = {9798400712487},
publisher = {Association for Computing Machinery},
address = {New York, NY, USA},
url = {https://doi.org/10.1145/3691620.3695065},
doi = {10.1145/3691620.3695065},
booktitle = {Proceedings of the 39th IEEE/ACM International Conference on Automated Software Engineering},
pages = {706–715},
numpages = {10},
location = {Sacramento, CA, USA},
series = {ASE '24}
}

@misc{transformers2025minilm,
	author       = { Sentence Transformers },
	title        = { all-MiniLM-L6-v2 (Revision c9745ed) },
	year         = 2025,
	url          = { https://huggingface.co/sentence-transformers/all-MiniLM-L6-v2 },
	publisher    = { Hugging Face }
}

@INPROCEEDINGS{he2017drain,
  author={He, Pinjia and Zhu, Jieming and Zheng, Zibin and Lyu, Michael R.},
  booktitle={2017 IEEE International Conference on Web Services (ICWS)}, 
  title={Drain: An Online Log Parsing Approach with Fixed Depth Tree}, 
  year={2017},
  volume={},
  number={},
  pages={33-40},
  doi={10.1109/ICWS.2017.13}}

@inproceedings{li2022circa,
author = {Li, Mingjie and Li, Zeyan and Yin, Kanglin and Nie, Xiaohui and Zhang, Wenchi and Sui, Kaixin and Pei, Dan},
title = {Causal Inference-Based Root Cause Analysis for Online Service Systems with Intervention Recognition},
year = {2022},
isbn = {9781450393850},
publisher = {Association for Computing Machinery},
address = {New York, NY, USA},
url = {https://doi.org/10.1145/3534678.3539041},
doi = {10.1145/3534678.3539041},
booktitle = {Proceedings of the 28th ACM SIGKDD Conference on Knowledge Discovery and Data Mining},
pages = {3230–3240},
numpages = {11},
location = {Washington DC, USA},
series = {KDD '22}
}

@inproceedings{li2022dejavu,
  author       = {Zeyan Li and Nengwen Zhao and Mingjie Li and Xianglin Lu and Lixin Wang and Dongdong Chang and Xiaohui Nie and Li Cao and Wenchi Zhang and Kaixin Sui and Yanhua Wang and Xu Du and Guoqing Duan and Dan Pei},
  title        = {Actionable and Interpretable Fault Localization for Recurring Failures in Online Service Systems},
  booktitle    = {Proceedings of the 30th ACM Joint European Software Engineering Conference and Symposium on the Foundations of Software Engineering (ESEC/FSE '22)},
  year         = {2022},
  pages        = {996--1008},
  publisher    = {ACM},
  doi          = {10.1145/3540250.3549092}
}

@article{wang2024aiopssurvey,
  author       = {Tingting Wang and Guilin Qi},
  title        = {A Comprehensive Survey on Root Cause Analysis in (Micro) Services: Methodologies, Challenges, and Trends},
  journal      = {arXiv preprint arXiv:2408.00803},
  year         = {2024},
  url          = {https://arxiv.org/abs/2408.00803}
}

@inproceedings{zhang2022deeptralog,
  author       = {Chenxi Zhang and Xin Peng and Chaofeng Sha and Ke Zhang and Zhenqing Fu and Xiya Wu and Qingwei Lin and Dongmei Zhang},
  title        = {{DeepTraLog}: Trace-Log Combined Microservice Anomaly Detection through Graph-based Deep Learning},
  booktitle    = {Proceedings of the 44th IEEE/ACM International Conference on Software Engineering (ICSE '22)},
  year         = {2022},
  pages        = {623--634},
  publisher    = {ACM},
  doi          = {10.1145/3510003.3510180}
}

@inproceedings{lee2023eadro,
  author       = {Cheryl Lee and Tianyi Yang and Zhuangbin Chen and Yuxin Su and Michael R. Lyu},
  title        = {{Eadro}: An End-to-End Troubleshooting Framework for Microservices on Multi-source Data},
  booktitle    = {Proceedings of the 45th IEEE/ACM International Conference on Software Engineering (ICSE '23)},
  year         = {2023},
  pages        = {1750--1762},
  publisher    = {IEEE},
  doi          = {10.1109/ICSE48619.2023.00150}
}

@inproceedings{zheng2024mulan,
  author       = {Lecheng Zheng and Zhengzhang Chen and Jingrui He and Haifeng Chen},
  title        = {{MULAN}: Multi-modal Causal Structure Learning and Root Cause Analysis for Microservice Systems},
  booktitle    = {Proceedings of the ACM Web Conference 2024 (WWW '24)},
  year         = {2024},
  pages        = {4107--4116},
  publisher    = {ACM},
  doi          = {10.1145/3589334.3645442}
}

@ARTICLE{guo2025automated,
  author={Guo, Xiansheng and Boateng, Gordon Owusu and Si, Haonan and Cao, Yu and Qiu, Yu and Lai, Zhexue and Li, Xilong and Liu, Xinhao and Chen, Cheng},
  journal={IEEE Communications Magazine}, 
  title={Automated Valet Parking and Charging: A Novel Collaborative AI-Empowered Architecture}, 
  year={2025},
  volume={63},
  number={1},
  pages={131-137},
  doi={10.1109/MCOM.001.2300824}}

@article{lu2025research,
  title={Research Progress in Multi-Domain and Cross-Domain AI Management and Control for Intelligent Electric Vehicles.},
  author={Lu, Dagang and Chen, Yu and Sun, Yan and Wei, Wenxuan and Ji, Shilin and Ruan, Hongshuo and Yi, Fengyan and Jia, Chunchun and Hu, Donghai and Tang, Kunpeng and others},
  journal={Energies (19961073)},
  volume={18},
  number={17},
  year={2025}
}

@inproceedings{weiss2026adselection,
  author    = {Matthias Wei{\ss} and Athreya Hosahalli Prakash and Maurice Artelt and Falk Dettinger and Nasser Jazdi and Michael Weyrich},
  title     = {Self-Adaptive Anomaly Detection with Reinforcement Learning and Human Feedback in Connected Vehicles},
  booktitle = {2026 IEEE 31st International Conference on Emerging Technologies and Factory Automation (ETFA)},
  year      = {2026},
  note      = {Accepted for publication}
}

@inproceedings{weiss2024simulating,
  author = "Weiß, Matthias and Stümpfle, Johannes and Dettinger, Falk and Jazdi, Nasser and Weyrich, Michael",
  title = "Simulating Cloud Environments of Connected Vehicles for Anomaly Detection",
  booktitle = "2024 Stuttgart International Symposium",
  publisher = "SAE International",
  month = jul,
  year = 2024,
  doi = "https://doi.org/10.4271/2024-01-2996",
  url = "https://doi.org/10.4271/2024-01-2996",
  issn = "0148-7191"
}

\end{document}